\documentclass[smallcondensed]{svjour3}                     
\smartqed  
\usepackage{graphicx}
\usepackage{mathptmx}      
\usepackage{latexsym}
\begin{document}

\title{Catastrophic senescence and semelparity \\ in the Penna aging model
}


\author{Chrysline Margus Pi\~nol         \and
        Ronald Banzon 
}


\institute{C.M.N. Pi\~nol \and R.S. Banzon \at
              Structure and Dynamics Group \\
              National Institute of Physics, University of the Philippines, \\
              Diliman, 1101 Quezon City, Philippines \\
              Tel.: +632-920-9749\\
              Fax: +632-928-0296\\
		  \email{cpinol@nip.upd.edu.ph, rbanzon@nip.upd.edu.ph}
           \and
           C.M.N. Pi\~nol \at
              Institute of Mathematical Sciences and Physics, University of the Philippines, \\
              Los Ba\~nos, 4031 Laguna, Philippines \\
}

\date{Received: 24 September 2010 / Accepted: 03 November 2010}

\maketitle

\begin{abstract}
The catastrophic senescence of the Pacific salmon is among the initial tests used to validate the Penna aging model. Based on the mutation accumulation theory, the sudden decrease in fitness following reproduction may be solely attributed to the semelparity of the species. In this work, we report other consequences of mutation accumulation. Contrary to earlier findings, such dramatic manifestation of aging depends not only on the choice of breeding strategy but also on the value of the reproduction age, $R$, and the mutation threshold, $T$. Senescence is catastrophic when $T \le R$. As the organism's tolerance for harmful genetic mutations increases, the aging process becomes more gradual. We observe senescence that is threshold dependent whenever $T>R$. That is, the sudden drop in survival rate occurs at age equal to the mutation threshold value.
\end{abstract}

\section{Introduction}
\label{intro}
Semelparous is the term given to species that undergo genetically programmed degeneration following procreation \cite{crespi}. They breed once at the same age, usually producing plenty of offspring, and die shortly after. These semelparous organisms include annual plants and many species of animals, particularly in the class Insecta \cite{coleman}. It is generally believed that these organisms allocate a huge amount of energy towards reproduction, thus making death inevitable \cite{smith}. Unlike in populations that breed repeatedly (iteroparous), senescence associated with semelparity or one-time reproduction is abrupt, and is often described as catastrophic. 

The life of a Pacific salmon, for example, traces a path from freshwater where they are born, to the ocean where they mature, and then back to freshwater where they spawn \cite{quinn}. Compared to other species which repeat this cycle, the Pacific salmon is only capable of one full roundtrip. They die in freshwater just after producing offspring. One probable cause for such behavior is starvation. Literature indicates that the salmon sometimes travels back more than 1200 kilometers to freshwater, generally without eating \cite{smdeoliveira04}. However, according to the mutation accumulation theory, the catastrophic senescence of the Pacific salmon is due solely to the semelparity of the species \cite{salmon}. That is, the drastic lowering of the salmon's fitness (to the point of death) immediately following procreation is completely explained by reproduction occurring only once at a fixed age. 

Another study \cite{meyer-ortmanns} demonstrated such dramatic aging manifestation utilizing a model that does not account for the effects of harmful mutations. Instead, some form of antagonistic pleiotropy was employed, promoting an increase in births by either shortening individual lifespan or delaying reproductive maturity.

This paper presents a further investigation on mutation accumulation and semelparity within the framework of the Penna model \cite{penna}, a simple, easily extensible tool for predicting many of the aging-related features found in biological systems \cite{smdeoliveira98}. Our results show aging trends other than the previously observed catastrophe.

\section{Methodology}
\label{method}
We implement the Penna model described in \cite{penna,penna-stauffer}. Individual characteristics are stored in 32-bit long genomes. For every year in the individual's life, one bit in the genome is read. Thus, only those genes located at bits less than or equal to the individual's current age are considered active. Zeroes correspond to healthy genes while ones indicate harmful mutations or diseases. The mutation threshold, $T$, defines the upper limit on the number of mutations individuals can undergo. An individual suffers a genetic death when the number of active mutations reaches this value. Reproductive maturity is achieved at age $R$. The number of offspring produced by a parent at each time of reproduction is given by the birth rate, $B$. Genetic traits, both active and inactive, are passed on from parent to offspring. During reproduction, $M$ new harmful, randomly located irreversible mutations are introduced into the newborn's genome. Usually, the mutation rate, $M$, is set to one \cite{malarz}. Environmental restrictions are taken into account by imposing an age-independent Verhulst factor, $V_t=1-N_t/K$. This corresponds to the probability that an individual survives given the current population size, $N_t$, and the carrying capacity, $K$. In simulations, the carrying capacity is commonly set to ten times the value of the initial population.

\paragraph{Simulation details} A population of $20000$ is propagated in an environment that can support a maximum of $200000$ individuals. Initially, all genomes are clean (no defective genes) and all individuals are at age zero. To consider semelparity, reproduction age is fixed at $R$. Because such species are generally known to produce plenty of offspring, the original paper utilized high birth rates. However, since the catastrophic behavior was shown to be independent of the birth rate \cite{salmon}, we limit its value to ten ($B=10$). Furthermore, we extend the demonstration to higher thresholds. 

The simulation runs for 3000 iterations. Simulated populations have already achieved steady state before this time. Demographic statistics such as the fraction of defective genes, Verhulst-related mortality and survival rates as functions of age are obtained for steady state populations associated with different parameter values. To minimize fluctuations, we consider mean population sizes of the last 300 iterations. Data hereafter are representative of ten independent simulation runs.

\section{Results and discussion}
\label{results}

\begin{figure}
  \includegraphics{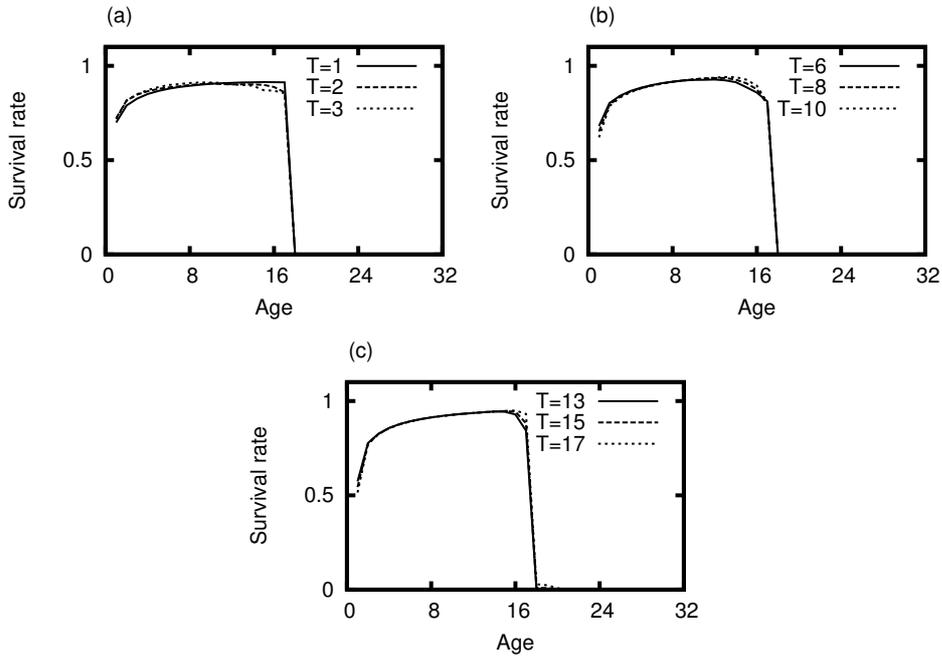}
\caption{Survival rates as a function of age in the case of reproduction happening only at $R=17$. The sharp decline in organism fitness occurs at $R=18$.}
\label{fig1}
\end{figure}

\begin{figure}
  \includegraphics{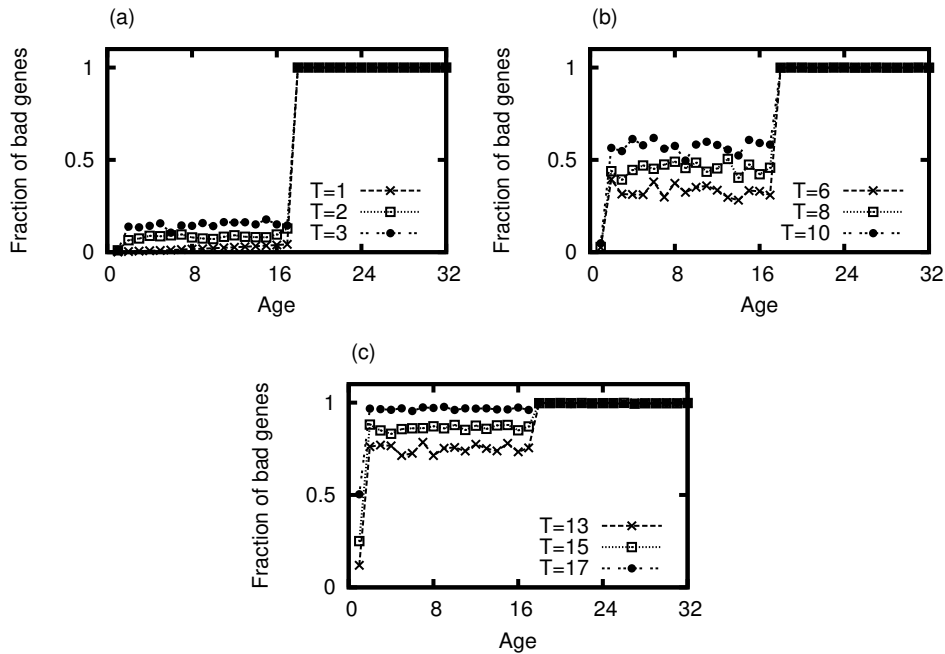}
\caption{Genetic profile of steady state populations associated with $R=17$. Harmful mutations accumulate at higher bits, starting at 18.}
\label{fig2}
\end{figure}

\begin{figure}
  \includegraphics{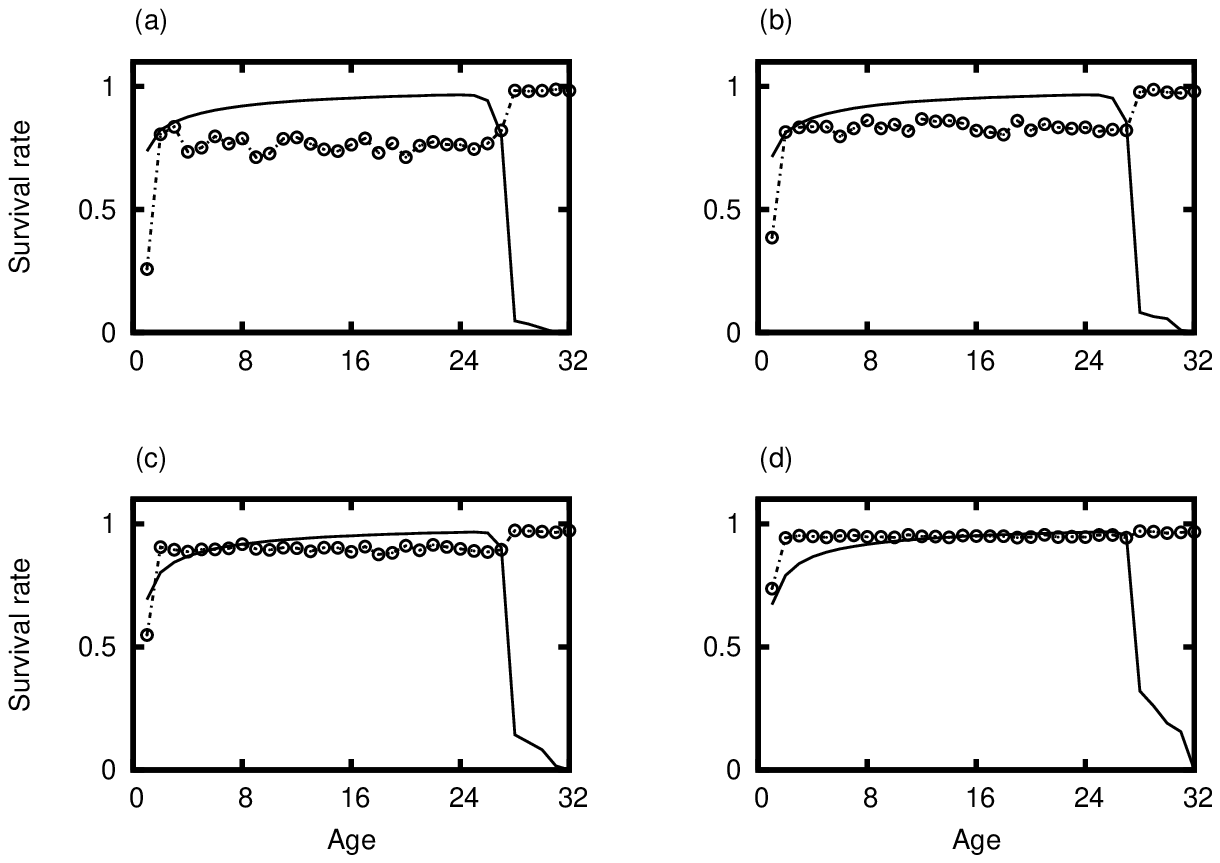}
\caption{Survival rates (solid lines) and genetic profile (or the fraction of defective genes - circles) of semelparous populations associated with $R=27$. Senescence is more gradual at higher threshold - (a) $T=21$, (b) $T=23$, (c) $T=25$, and (d) $T=27$.}
\label{fig3}
\end{figure}

Aging is said to be catastrophic for populations that maintain a high level of fitness and condition until shortly before death \cite{ricklefs}. Losses in functional capabilities directly affect an organism's ability to withstand a given environment. Thus, here, we relate fitness to the survival rate - the ratio between the population with age $a$ at time $t$ and the population with age $a-1$ of the previous iteration, $N_a(t)/N_{a-1}(t-1)$.

\subsection{Catastrophic senescence}
\label{cat}

The survival rate of semelparous species usually goes down, from some relatively high value to zero, at the age immediately following reproduction. For populations associated with $R=17$ and $T \le R$, the sudden decrease in organism fitness happens at $R=18$ regardless of the value of $T$ (Fig.~\ref{fig1}).

Catastrophic senescence is closely associated with the accumulation of bad mutations at the end part of the genome \cite{martins-mossdeoliveira}. The life cycle of semelparous species can be viewed as a two-phase process driven by different mechanisms \cite{vaupel}. It starts with a juvenile non-reproductive phase ($age < R$), followed by the adult phase which is characterized generally by reproductive maturity. Genes switched on during the juvenile phase are called housekeeping genes, while those activated in the adult phase are referred to as death genes \cite{niewczas}. The concept of mutation accumulation introduces strong selection pressure during the non-reproductive phase in order to regulate harmful mutations that are passed on from parent to offspring \cite{smdeoliveira04}. Genetic housekeeping works to minimize harmful mutations expressed prior to reproduction. Selection, thus, acts to keep the first part of the genome clean. When $T=1$, only individuals with no expressed mutations at bits less than or equal to $R$ are allowed to procreate. The model is designed, however, in such a way that inactive genes are also inherited. The selection process brought forth by the dynamics of mutation accumulation does not screen bad mutations that are activated only after reproduction. As a result, diseases pile up at the higher bits. Fig.~\ref{fig2} shows the corresponding genetic profile of the steady state populations in Fig.~\ref{fig1}. The fraction of deleterious mutations occurring at bits greater than $R$ is 1.0. Moreover, we see that when $T \le R$ the survival rates fall off at the age when the fraction of defective genes is equal to 1.0 (cf. Figs.~\ref{fig1} and ~\ref{fig2}). Thus, the catastrophic loss in organism fitness is a consequence of the high density of bad mutations switched on immediately after reproduction.

The strength of selection diminishes with increasing mutation threshold. At higher $T$, more mutations are preserved in the gene pool. Such genes are believed to be essential for the organism's growth. It was previously demonstrated \cite{salmon} that although higher threshold values yield larger populations, a different $T$ does not alter the survival rate. On the contrary, Fig.~\ref{fig3} shows that as $T$ is increased further, the manifestation of aging becomes more gradual. The abrupt decline in organism fitness still occurs at $R+1$. However, a fraction of the population is observed to survive after procreation.

\subsection{Threshold dependence}
\label{tdep}

\begin{figure}
  \includegraphics{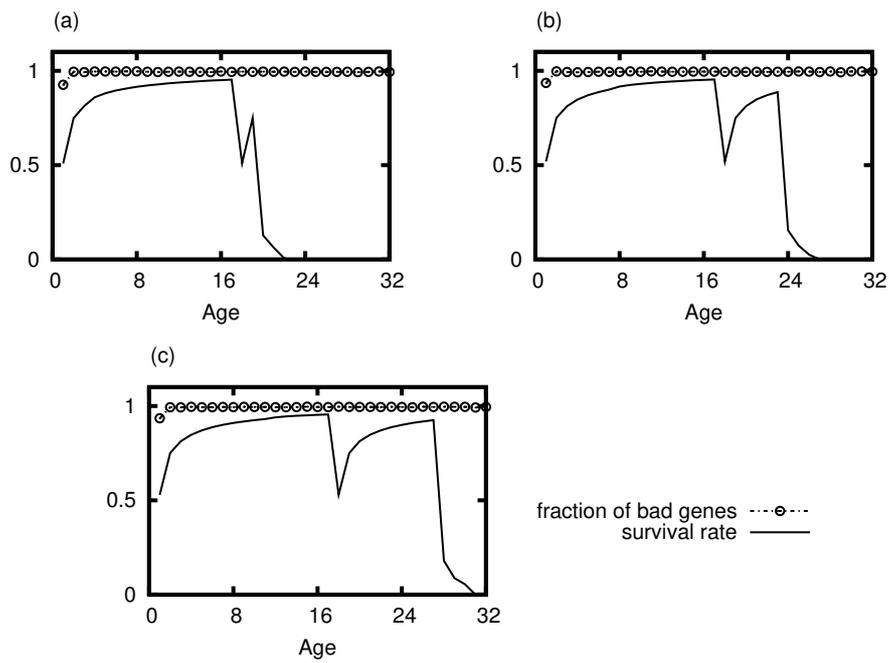}
\caption{Behavior of the survival rates  when $T>R$. Here, $R=17$: (a) $T=20$, (b) $T=24$, and (c) $T=28$.}
\label{fig4}
\end{figure}

\begin{figure}
  \includegraphics{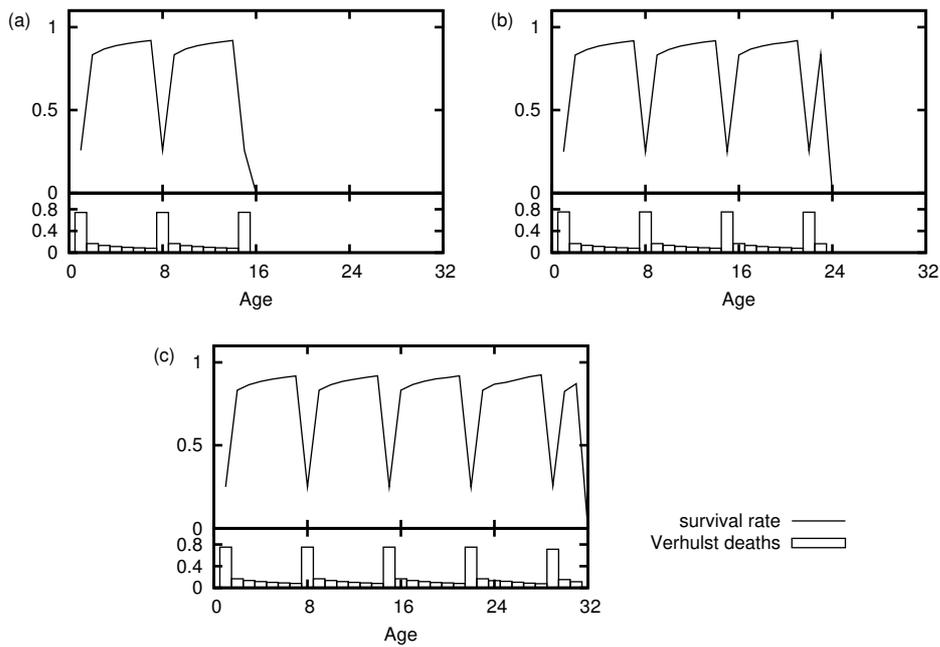}
\caption{Survival rate (top) and Verhulst mortality rate (bottom) of steady state populations associated with $R=7$: (a) $T=16$, (b) $T=24$ and (c) $T=32$.}
\label{fig5}
\end{figure}

When $T>R$, selection is very weak. Genetic housekeeping is perfomed solely by the Verhulst factor because threshold deaths do not occur until age $T$. In contrast to mutation accumulation, the Verhulst factor acts at random. It is incapable of discriminating between good and bad genomes, killing individuals regardless of fitness. It does very little in preventing the spread of diseases among the bits. Unlike in the previous case where bad mutations are more concentrated at higher bits, we observe a high density of deleterious genes occurring even at lower bits. In Fig.~\ref{fig4}, the fraction of defective genes is approximately 1.0 for all bits. For these populations, the drop in the survival rate is manifested as soon as organisms reach $age=T$. Senescence, in this case, is threshold dependent.

Plots in Fig.~\ref{fig4} also register a local minimum for the survival rates at $R+1$. The frequency of such dips increases with lower $R$ and higher $T$. Fig.~\ref{fig5} presents the survival rates of populations associated with $R=7, T=16,24$ and 32. The corresponding bottom plots show Verhulst-related death rates. Notice that the dips happen at ages when the Verhulst mortality rate is highest. Recall that in our simulations, we begin with a set of individuals characterized by the same age and genome. The resulting population is, thus, structured in such a way that those with ages separated by a factor $R$ are clustered together. The Penna model follows Gompertz law \cite{smdeoliveira98,puhl} which implies that most individuals are newborns. Because the Verhulst killing effect increases with total population size, it is felt most by those grouped with age 1. The observed periodicity, hence, arises from our choice of initialization. This is not seen in other Penna model implementations whose initial population is characterized by random ages and genomes.

\begin{figure}
  \includegraphics{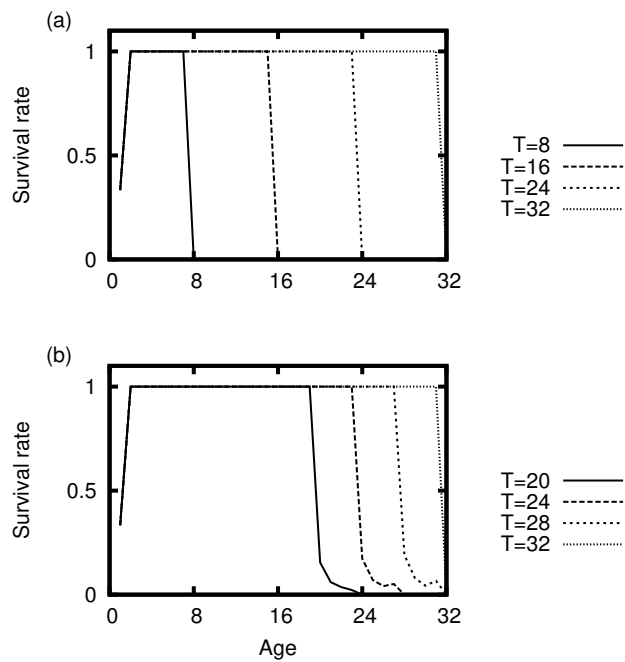}
\caption{Survival rate resulting from a VB implementation: (a) $R=7$ and (b) $R=17$. In these runs, the birth rate was reduced significantly to avoid extinction via overpopulation ($B=3$).}
\label{fig6}
\end{figure}

We lose the dips by limiting Verhulst deaths to newborns only - VB implementation \cite{dabkowski}, as in Fig.~\ref{fig6}. Note, however, that in our demonstration, we had to use a lower birth rate, $B=3$, in order to avoid extinction via overpopulation. Higher $B$ values introduce sudden fluctuations in the total population (beyond the set carrying capacity) which kills all newborns and, eventually, the entire population. The maximum lifespan is generally longer for the VB population \cite{martins}. For the above presented semelparous cases, this is evident only at larger $R$ values (cf. Figs.~\ref{fig4},~\ref{fig5}, and ~\ref{fig6}). 

\section{Conclusion}
The theory of mutation accumulation depicts senescence as a consequence of increased mutation load occurring at later ages \cite{partridge-barton}. Within the framework of the Penna model, this is seen in the high concentration of bad genes located at bits greater than $R$. An increase in the tolerance for harmful mutations weakens selection and increases the probability of finding a deleterious mutation at a lower bit. When $T>R$, this probability is equivalent to unity, for all bits. Ignoring Verhulst effects, the corresponding survival rates are henceforth flat at 1.0 from ages 1 to $T-1$. Such lack of age-specific variation in fitness and mortality rate has been reportedly observed in coldwater fish, bivalves, turtles, whales, and naked mole-rats \cite{finch,buffenstein}.

Aging manifests differently depending on the choice of reproductive strategy. Senescence associated with semelparity is generally catastrophic, while that with iteroparity (or multiple breeding) is gradual. This characteristic makes organisms that exhibit interspecific variations in parity very useful in evolutionary studies \cite{crespi}. The catastrophic behavior of the Pacific salmon has been modeled extensively with tools that are based on different aging theories. Previous works \cite{salmon,meyer-ortmanns} suggest that the one-time reproductive strategy is the only ingredient needed to explain the substantial loss in functional abilities immediately following procreation. Utilizing a model that is based on the mutation accumulation theory, we were able to find other limiting factors that yield catastrophic senescence. When $T \le R$, the high density of bad mutations occurring at higher bits causes an abrupt decline in the survival rate at age $R+1$. On the other hand, when $T>R$, aging is catastrophic at age $T$ (threshold dependent). The major drop in fitness happens at age $T$, when threshold deaths begin to take place.

\end{document}